\documentclass[12pt]{article}
\usepackage{amsmath,amssymb,amsthm,color}
\usepackage{graphicx}
\usepackage{cite}
\usepackage{epsfig}
\renewcommand{\baselinestretch}{2}

\begin{document}
%
\title{Diversified properties of carbon substitutions in silicene \\}
\author{
\small Hai-Duong Pham$^{a}$, Shih-Yang Lin$^{a}$, Godfrey Gumbs$^{b}$, Nguyen Duy Khanh$^{c}$, Ming-Fa Lin$^{a,*}$ $$\\
\small  $^a$Department of Physics, National Cheng Kung University, Tainan 701, Taiwan\\
\small  $^b$Department of Physics and Astronomy, Hunter College of the City University of New York,\\
\small 695 Park Avenue, New York, New York 10065, USA\\
\small  $^c$Advance Institute of Materials Science, Ton Duc Thang University, 19 Nguyen Huu Tho Street,\\
\small District 7, Ho Chi Minh City, Vietnam \\
 }
\renewcommand{\baselinestretch}{1}
\maketitle

\renewcommand{\baselinestretch}{1.4}
\begin{abstract}

The theoretical framework, which is built from the first-principles results, is successfully developed for investigating emergent two-dimensional (2D) materials, as it is clearly illustrated by carbon substitution in silicene. Computer coding with the aid of VASP in conjunction with data analysis from the multi-orbital hybridizations [spin configurationsf ] are thoroughly identified from the optimal honeycomb lattices, the atom-dominated energy spectra, and the spatial charge density distributions. The atom and orbital-decomposed van Hove singularities [the net magnetic moments], being very sensitive to the concentration and arrangements of guest atoms. All the binary 2D silicon-carbon compounds belong to the finite- or zero-gap semiconductors, corresponding to the thoroughly/strongly/slightly modified Dirac-cone structures near the Fermi level. Additionally, there are frequent $\pi$ and $\sigma$ band crossings, but less anti-crossing behaviors. Apparently, our results indicate the well-defined $\pi$ and $\sigma$ bondings.

\vskip 1.0 truecm
\par\noindent

\par\noindent  * Corresponding author. {~ Tel:~ +886-6-2757575-65272}\\~{{\it E-mail address}: mflin@mail.ncku.edu.tw (M.F. Lin)}
\end{abstract}

\pagebreak
\renewcommand{\baselinestretch}{2}
\newpage

{\bf 1. Introduction}
\vskip 0.3 truecm

Chemical substitutions on layered materials are capable of band structure tailoring which could lead to significant modifications of the properties of pristine lattices through very strong host-guest multi-orbital hybridizations. With the use of modern experimental growth techniques,  ternary and binary compounds, which are characterized by B$_x$C$_y$N$_z$, have been successfully synthesized for three-dimensional (3D) bulk systems \cite{9-1,9-2}, two-dimensional (2D) layers \cite{9-6}, one-dimensional (1D) cylindrical nanotubes \cite{9-9}, 1D nanoribbons \cite{9-12}, and zero-dimensional (0D) quantum dots \cite{9-15}. Their geometric structures vary from three to zero dimensions, as observed in carbon-related systems \cite{9-17,9-18}. This clearly indicates that each atom possesses at least three half-filled orbitals.  Similar syntheses have been performed for the high-potential Si-C compounds \cite{9-21}. In general, it would be routine to produce the above-mentioned compounds, whereas the opposite might be true for specific components. For example, using high-performance experimental techniques, it may be difficult to manipulate the ratio between the [B, C, N]/[Si, C] atoms. Such non-monoelement condensed-matter systems have been predicted or found to exhibit the observable energy gaps, or belong to specific semiconductors. The main reason lies in the distinct ionization energies  of their components, being consistent with the tight-binding mode for non-vanishing diagonal Hamiltonian matrix elements [the sublattice-dependent site energies]. Gap engineering could be achieved by transforming the strength relations in the $\pi$, $\sigma$ and sp$^3$ bondings the [p$_z$-, (s, p$_x$, p$_y$)- and (s, p$_x$, p$_y$, p$_z$)-orbital hybridizations].

\medskip
\par

Chemisorption and substitution of silicon atoms on graphene are two very interesting procedures, since first-principles predictions are available for understanding the important differences between these two types of chemical modification \cite{9-22,9-23}, and providing additional information regarding another approach for the formation of silicon-carbide compounds \cite{9-24}. Based on the viewpoint of a carbon-created honeycomb lattice,  silicon atoms are regarded as adatoms and guest ones, respectively, in these two cases. The former presents optimal positions at the bridge sides above the graphene surface \cite{9-25}, while the latter shows the [Si, C]-co-dominated A and B sublattices with deformed  hexagons \cite{9-26}. Both silicon and carbon atoms have rather active dangling bonds, leadings to significant multi-orbital hybridizations in Si-C [Si-Si and C-C] bonds. Most important, the planar bondings between Si and C atoms are expected to be stronger and complicated, compared with the perpendicular configuration. That is to say, the $\pi$ and $\sigma$ [$\pi$] bondings of graphene will be significantly modified by Si-substitutions \cite{9-27}. This is directly reflected in the diversified properties, the spatial charge distributions, atom-dominated energy bands, and atom- as well as orbital-projected density-of-states. Furthermore, their gradual transformations could be achieved through the Si-/C-substituted graphene/silicene honeycomb lattice.

\medskip
\par

The present work is focused on the diverse geometric, electronic and magnetic properties of C-substituted silicene. A theoretical framework, which is based on  multi-orbital hybridizations and atom-created spin configurations, is further developed to clearly analyze the concentration- and configuration-dependent phenomena. For example, the zero- and finite-gap behaviors, the main features of $\pi$ and $\sigma$ bands could be understood from the specific relations among the hybridized $\pi$, $\sigma$ and sp$^3$ chemical bondings. On the experimental side, a controllable synthesis way, using MTMS/hexane as precursors, is proposed to generate the large-area graphene-based Si-C binary 2D compounds \cite{9-52}. Furthermore, the silicon-carbide nanosheets are successfully synthesized  by a catalyst free carbothermal method and post-sonication process  \cite{9-53}, in which the AFM measurements show the average  thickness of ${\sim\,2-3}$ nm and size of ${\sim\,2}$ $\mu$m.

\vskip 0.6 truecm
\par\noindent
{\bf 2. Method of calculation }
\vskip 0.3 truecm

Our investigation of the diverse properties of carbon-substituted silicene is base on density functional theory using VASP codes \cite{method1,method2}. The exchange and correlation energies due to many-particle Coulomb interactions were calculated with the use of the Perdew-Burke-Ernzerhof (PBE) functional under the generalized gradient approximation \cite{method3}, whereas the electron-ion interactions can be characterized by the projector augmented wave (PAW) pseudopotentials \cite{method4}. A plane-wave basis set with a maximum kinetic energy cutoff of 500 eV was chosen to expand the wave function. In a direction perpendicular to the silicene plane, a vacuum layer with a thickness of 15 $\AA$ was added to avoid interactions between adjacent unit cells. The k-point mesh was set as $9\times 9 \times 1$ in geometry optimization, $100\times 100\times 1$ for further calculations on electronic properties via the Monkhorst-Pack scheme. During the ionic relaxations, the maximum Hellmann-Feynman force acting on each atom is less than $0.01$ eV/Å  whereas the convergent energy scale was chosen as  $10-5$ eV between two consecutive steps.

\vskip 0.6 truecm
\par\noindent
{\bf 3. Numerical Results and discussion}
\vskip 0.3 truecm

\subsection{Geometric structure of carbon-substituted silicene}

\medskip
\par

Carbon-substituted silicon systems are capaable of possessing unusual geometries, as it is clearly indicated in Table 1 and Figs. 2(a) through 2(l). Four types of typical C-substitution configurations, which cover meta, ortho, para, and single cases, are chosen for a model investigation. In general, the third and fourth types, respectively, possess the lowest and highest ground state energies, i. e., the para-configuration is the most stable among them, or it is expected to be relatively easily synthesized in experimental growths. Carbon and silicon atoms, respectively, possess four outer orbitals of [2p$_x$, 2p$_y$, 2p$_z$, 2s] and  [3p$_x$, 3p$_y$, 3p$_z$, 3s]. In addition to the pure ones, the substitutions of the former, as clearly illustrated in Fig. 2(a)  through 2(l), will create the hybridized $\pi$ and $\sigma$ bondings [2p$_z$-3p$_z$ and (2s, 2p$_x$, 2p$_y$)-(3s, 3p$_x$, 3p$_y$)] in silicene honeycomb lattice [the standard orbital hybridizations in Fig. 2(a)].  This seems to be responsible for the extremely non-uniform chemical/physical environments in an enlarged unit cell. Consequently, the Si-Si/Si-C bond lengths and the height difference between A and B sublattices might lie in specific ranges [Table 1]. Specifically, the full C-substitution case, with the only uniform environment, is clearly different from the pristine one, since they, respectively, have the planar and buckled honeycomb lattices [Figs. 2(b) and 2(a); $\delta_z$=0 $\sim\,0.48$ $\AA$]. When the C-concentration is sufficiently low, the buckled structures will recover, such as, the C:Si ratio lower than 15$\%$. For the other conditions, the existence of buckling strongly depends on the concentration and configuration of the substitution.

\begin{figure}
\centering
\includegraphics[width=16.2cm, height=10cm]{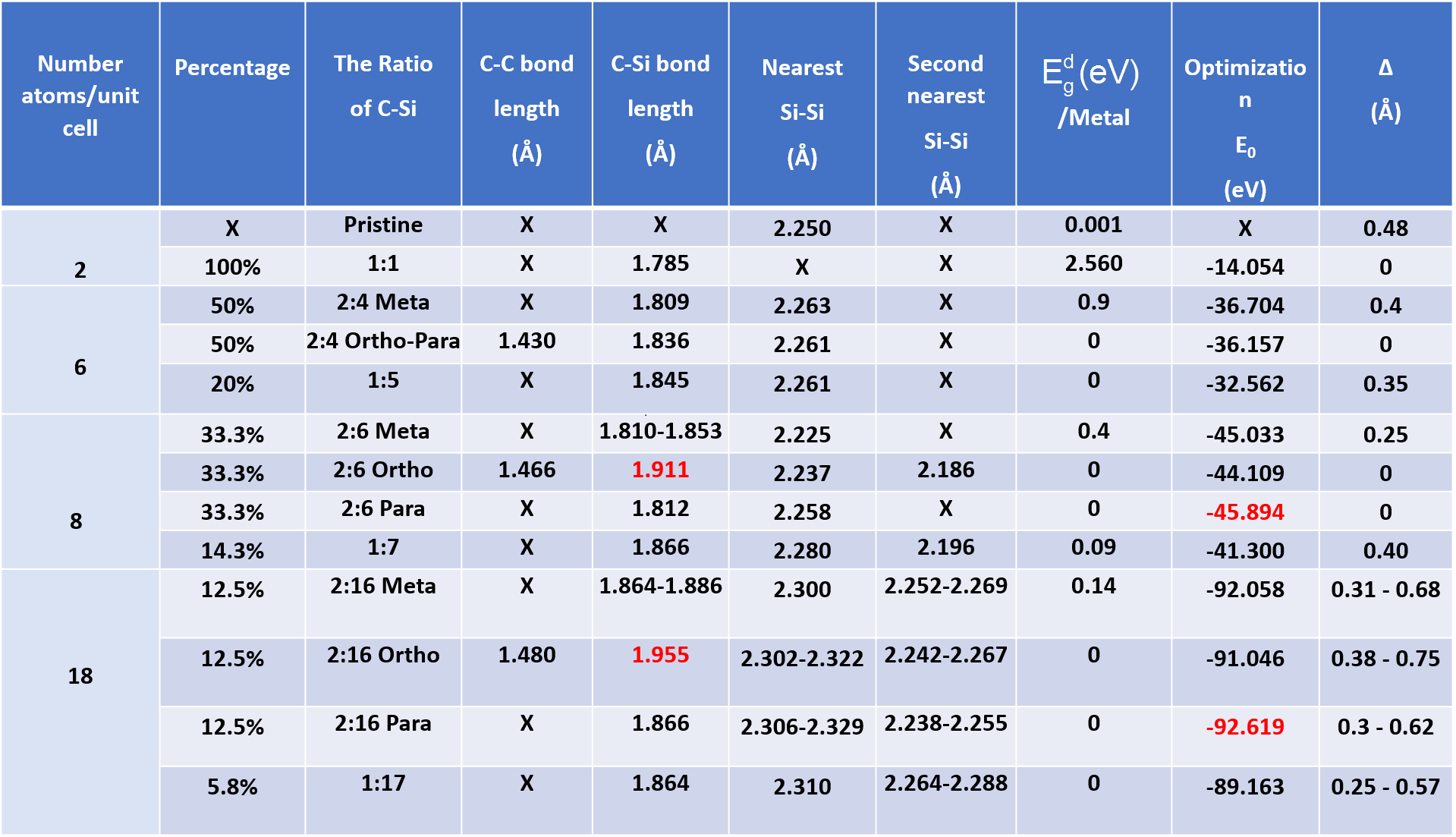}
\caption{Table: The optimal geometric structures of carbon-substituted silicene systems under various concentrations and configurations with the C-C, C-Si and Si-Si bond lengths, the band gaps, ground state energies  per unit cell, together with the height differences between A and B sublattices.}
\end{figure}

\begin{figure}
\centering
\includegraphics[width=12cm, height=18cm]{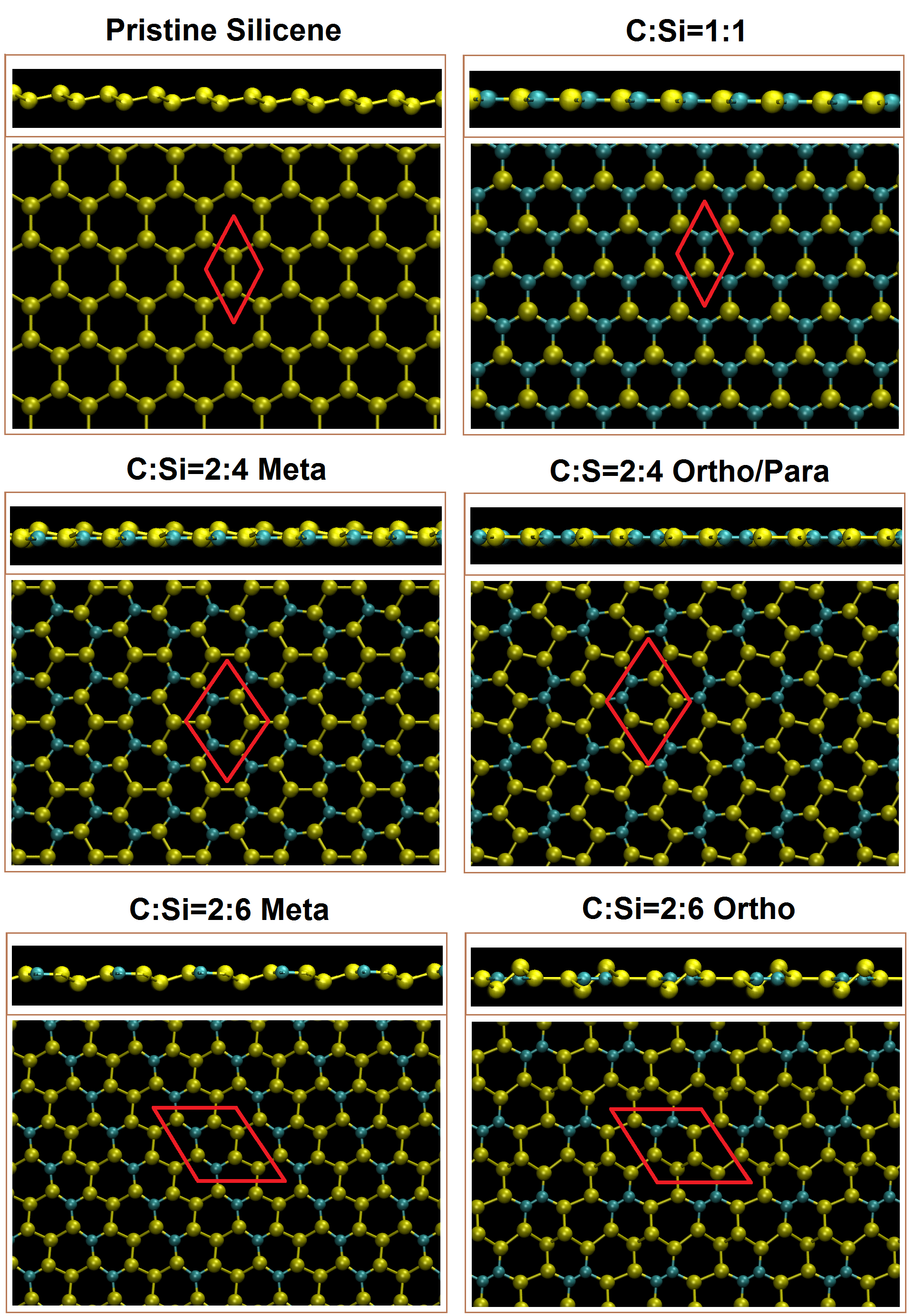}
\label{fgr:1}
\end{figure}
\begin{figure}
\centering
  \includegraphics[width=12cm, height=18cm]{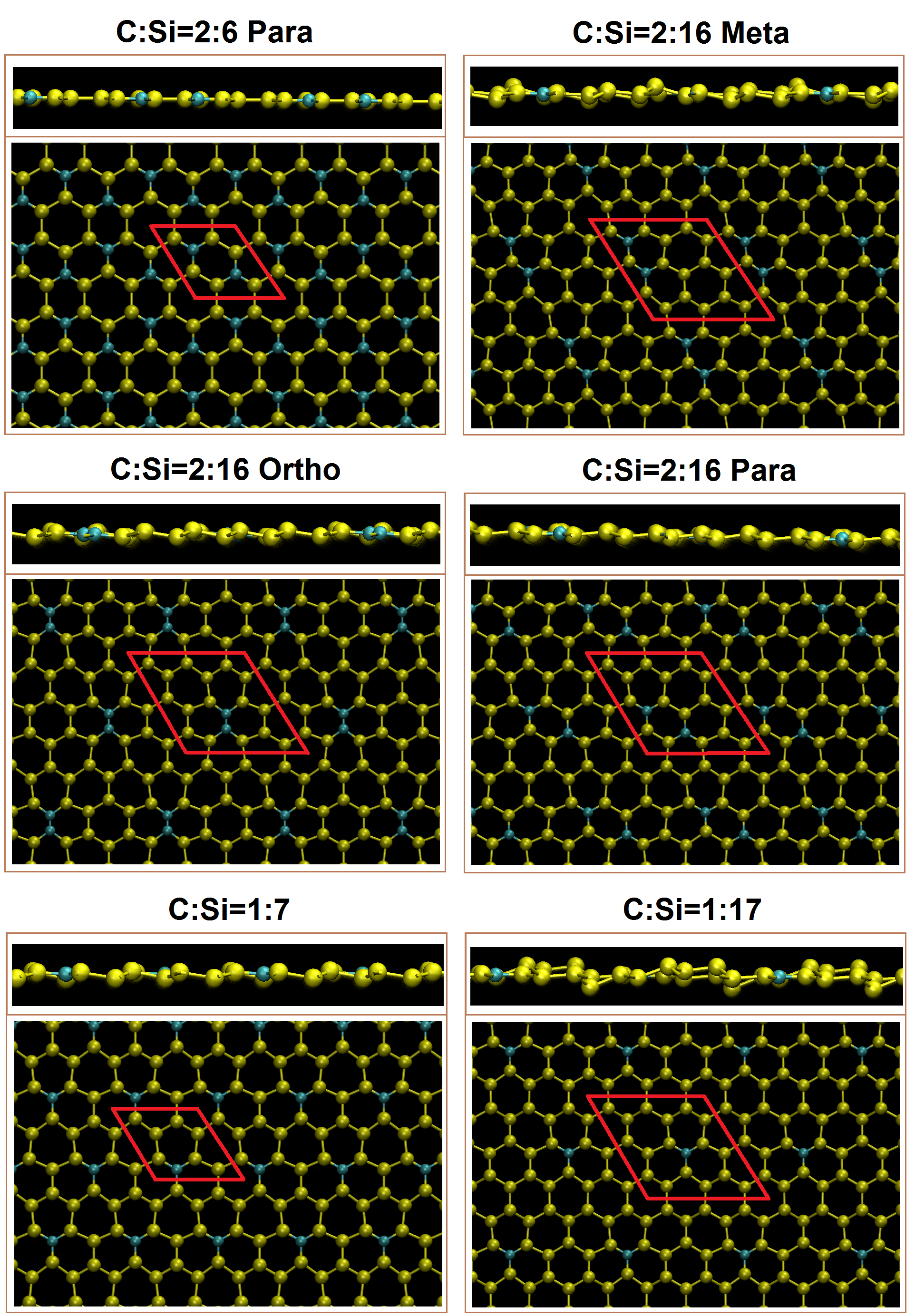}
\caption{The top-view optimal geometric structures for C-substituted silicene systems; the (a) pristine, (b) full-, (c) 50$\%$-meta-, (d) $50$\%-ortho-/para-, (e) 33.3$\%$-meta, (f) 33$\%$-ortho,  (g) 3 3$\%$-para- (h) 12.5$\%$-meta-, (i) 12.5$\%$-ortho-, (j)12.5$\%$-para. (k) 14.3$\%$-single- and (l) 5.8$\%$-single-substitution cases. Also shown in the narrow windows are the side views for the buckled structures.}
  \label{fgr:2}
  \end{figure}

\subsection{Band structure tailoring of Carbon-substituted silicene systems}

\medskip
\par
The specific relations among the $\pi$, $\sigma$ and ${sp^3}$ bondings in a pristine monolayer silicene account for the main features of the band structure. The electron-hole symmetry around the Fermi level is only weakly modified at low energy, as illustrated in Fig. 3(a). The first pair of valence and conduction bands. being nearest to ${E_F=0}$, are initiated from the stable K/K$^\prime$ valley. A very narrow band gap of ${E_g\sim\,0.01}$ meV comes to exist between  slightly separated   Dirac cones, mainly due to the weak spin-orbital coupling [a single-particle interaction]. This result is consistent with that obtained using the tight-binding model \cite{9-57}.  Additionally, they show the valence/conduction saddle M-point structure at -1.02 eV/0.59 eV. Finally, the $\pi$ band is ended at the stable $\Gamma$ valley, in which the $\pi$-band energy width for valence states, the energy spacing between the initial K point and the final $\Gamma$ point is about 3.2 eV.  Apparently, such electronic spectrum originates from the $\pi$ bonding of pure 3${p_z}$-3${p_z}$ orbital hybridizations in the buckled honeycomb lattice. On the other hand, the four-fold degenerate $\sigma$ bands, which arise  mainly from [3p$_x$, 3$p_y$]-orbital bondings, obey parabolic energy dispersion relations from the stable $\Gamma$ valley at ${E^v\sim\,-1.04}$ eV.  The $\sigma$ and $\pi$ valence bands have the obvious or observable mixing behavior along the $\Gamma$K and $\Gamma$M directions, since there exists a weak, but significant sp$^3$ bonding. Additionally, the band width of the first $\sigma$-electronic states along $\Gamma$MK is about 3.45 eV. Generally, all $\pi$- and $\sigma$-bands [three-band] widths could be well defined for monolayer silicene. In addition, all ${\bf k}$-states in the energy bands are doubly degenerate in terms of spin degree of freedom, since the spin-up- and spin-down-dominated states have identical energy spectra.

\begin{figure}
\centering
\includegraphics[width=15cm, height=18cm]{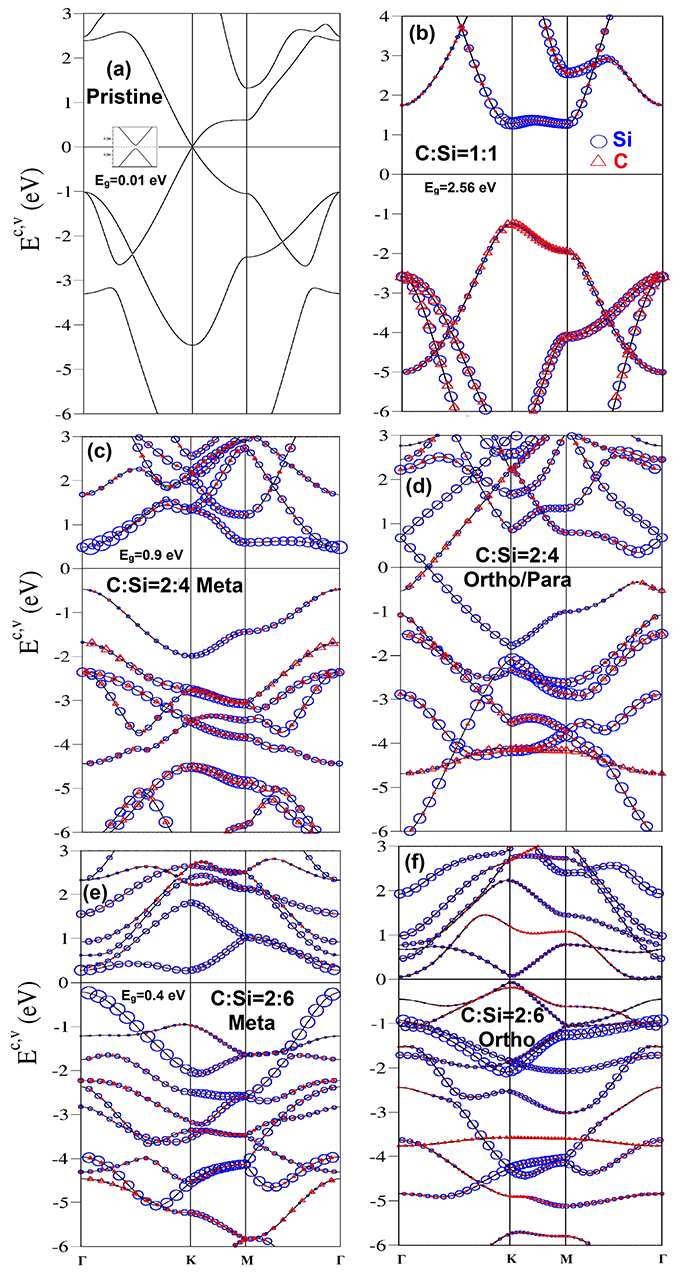}
\label{fgr:3}
\end{figure}

\begin{figure}
\centering
\includegraphics[width=15cm, height=18cm]{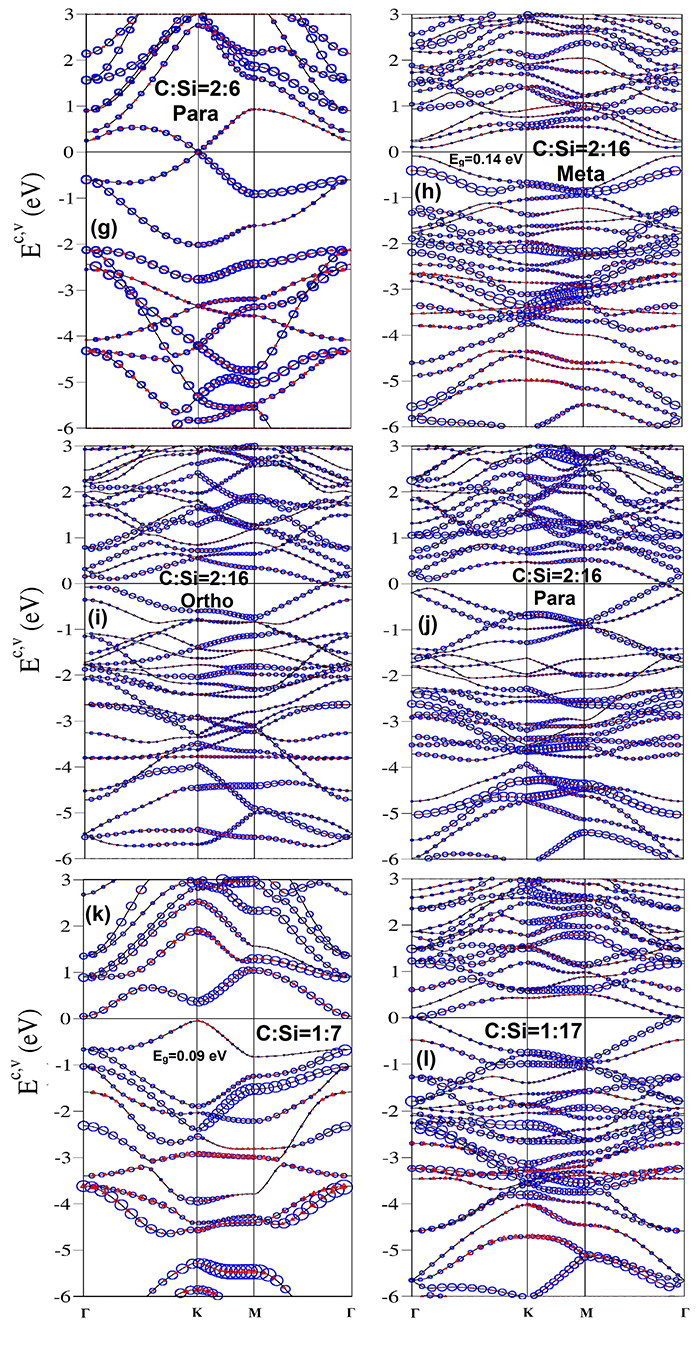}
\caption{Electronic structures, with the dominance circles of silicon and carbon atoms [the blue and red ones], for the  C-substituted silicene systems; the (a) pristine, (b) [1:1]-, (c) [2:4]-meta-, (d) [2:4]-ortho-/para-, (e) [2:6]-meta-, (f) [2;6]-ortho-, (g) [2:6]-para-, (h) [2:16]-meta-, (i) [2:16]-ortho-, (j) [2:16]-para-, (k) [1:7]-single-, (l) [1:17]-para substitution configurations.}
\label{fgr:4}
\end{figure}

\medskip
\par

 The full carbon-substitution silicene system, i.e., the silicon-carbon binary compound \cite{9-58}, exhibits an unusual electronic energy spectrum as shown in Fig. 3(b), being totally different from the pristine case [Fig. 3(a)]. The occupied valence bands are highly asymmetric to the unoccupied conduction bands near ${E_F=0}$.  Such material  is classified as a direct-gap insulator with a band gap larger than 2.56 eV at the K/K$^\prime$ point, as observed in a wide-gap monolayer boron-nitride system \cite{9-59}. This is closely related to the strongly modified Dirac-cone structure across the Fermi level with highly anisotropic enegy dispersions [e.g., the partially flat conduction band along KM]. Both valence and conduction bands nearest to $E_F$ are, respectively, $\pi$ and $\pi^\ast$ bands due to the  ${2p_z}$-${3p_z}$ impure orbital hybridizations in Si-C bonds. The higher/lower ionization energy of 2p$_z$/3p$_z$ orbitals further leads to the creation of a band gap for the honeycomb lattice,  and the C-/Si-dominance of the $\pi$/${\pi^{\ast}}$ bands [the red and blue circles].  The first pair of energy bands are highly anisotropic. For example, the valence/conduction $\pi$/${\pi^\ast}$ states, which are initiated from the K valley, show different group velocities along the KM and K$\Gamma$ directions, especially  for the saddle point/the partially flat dispersion related to the M point. Furthermore, the $\pi$-band energy width is ${E_w\sim\,3.71}$ eV] according to the difference between the K- and $\Gamma$-point energies [${-1.50}$ eV and ${-5.20}$ eV. Compared with a pristine 3p$_z$-3p$_z$ case [${E_w\sim\,3.20}$ eV in Fig. 3(a)], the larger band width might indicate the stronger 2p$_z$-3p$_z$  bonding. The two $\sigma$ bands, which arise from [3p$_x$, 3p$_y$]-[2p$_x$, 2p$_y$] hybridizations, appear at ${E^v\sim\,-2.60}$ eV initiated from the $\Gamma$ point, but remain fourfold degenerate electronic states there. Furthermore, the contributions from the different orbitals are comparable in the [C, Si]-co-dominated energy bands. The concave-downward parabolic valley near the $\Gamma$ point is dramatically transformed into cancave-upward ones/the saddle-point forms along the $\Gamma$K/$\Gamma$M direction. Consequently, the first $\sigma$-bandwidth is approximately $4.30$ eV. There exist direct crossings of the $\pi$ and $\sigma$ valence bands along any direction, e.g., M$\Gamma$ and K$\Gamma$. This result clearly reveals no evidence of sp$^3$ bonding \cite{9-60}. As a result, the fundamental properties of the [1:1] Si-C compound are dominated by the well-behaved $\pi$ and $\sigma$ bondings in the absence of sp$^3$ ones.


\begin{figure}
\centering
\includegraphics[width=15cm, height=12 cm]{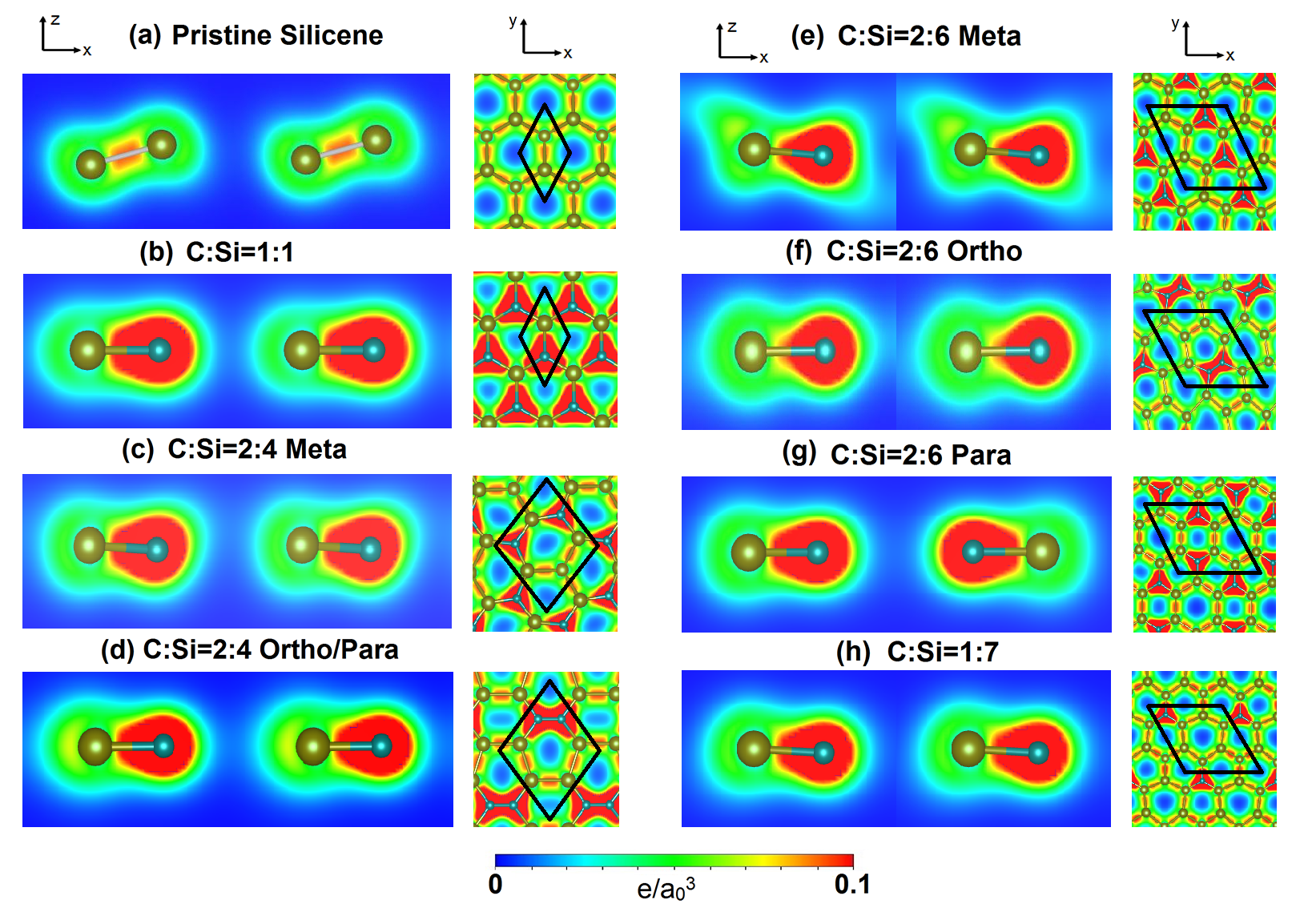}
\caption{The spatial charge densities of the  carbon-substituted silicene materials for the various concentrations and configurations: (a) the absence of guest atoms, (b) [1:1], (c) [2:4]-meta, (d) [2:4]-ortho-/para, (e) [2:6]-meta, (f) [2;6]-ortho, (g) [2:6]-para and (h) [1:7] conditions. They are shown on the ${[x, z]}$ and ${[x, y]}$ planes, clearly illustrating the $\pi$ and $\sigma$ chemical bondings and their orthogonality/non-orthogonality.}
\label{fgr:8}
\end{figure}

\medskip
\par

Electronic energy spectra are greatly enriched through the modulations of concentration and distribution configuration. The [2:4]-meta and [2:4]-ortho/para cases, as clearly displayed in Figs. 3(c) and 3(d), present finite- and zero-gap behaviors, respectively, corresponding to the highest occupied state and lowest one at the $\Gamma$ point [${E_g\sim\,0.90}$ eV] and the gapless Dirac-cone structure along the $\Gamma$K direction [but not $\Gamma$M]. That the low-lying valence and conduction bands are initiated from the stable $\Gamma$ valley is associated with the zone-folding effect.  According to the pristine and full-substitution configurations [Figs. 3(a) and 3(b)], the valence and conduction bands in the energy range of ${|E^{c,v}<1.30|}$ eV, including the first pair, mainly arise from the [2p$_z$, 3p$_z$]-orbital hybridizations. Such bondings are responsible for the low-energy physical properties. This result is also confirmed by the orbital-projected density-of-states [discussed later in Fig. 5(b)]. It should be noted that the latter [Fig. 3(d)] belongs to a zero-gap semiconductor because of the vanishing density-of- states at the Fermi level.  The low-lying energy bands are dominated by the 3p$_z$ orbitals of silicon-host atoms because of the higher weight. More energy subbands come to exist under the enlarged unit cells. Therefore, the band crossings and anti-crossings would happen frequently. In addition, it might be able to define the $\pi$-band energy widths through the $\Gamma$K$\to $KM$\to $M$\Gamma$ direction [examinations from the 3p$_z$-projected density-of-states in Figs. 5(c) and 5(d)], respectively, corresponding to $4.01$ eV and $4.70$ eV for the meta- and ortho-configurations. It would be very difficultm or even meaningless, in characterizing band widths  with further decrease of guest-atom concentration, as a result of more complicated valence subbbands. Concerning the first $\sigma$ valence bands, they are roughly identified from the initial $\Gamma$-states at ${E^v\sim\,-1.80}$ eV for the meta-case [Fig. 3(c)], while the very strong zone-folding effects forbid their characterizations under the ortho-condition [Fig. 3(d)].

\medskip
\par
 Very interestingly, two kinds of band properties are also revealed in the [2:6] cases, as clearly indicated in Figs. 3(e)-3(g). The meta-, ortho- and para-configurations, respectively, exhibit the small-, narrow- and zero-gap behaviors [the finite- and zero-gap semiconductors] according to the first pair of energy bands nearest to the Fermi level. The first case has an indirect band gap of ${E_g\sim\,0.31}$ eV, being determined by the specific energy spacing between the highest occupied state at the K point and the lowest unoccupied state at the $\Gamma$ point [Fig. 3(e)]. Furthermore, the first valence band has an oscillatory energy dispersions.  As a result, two stable valleys dominate low-energy physical phenomena simultaneously. The significant overlaps of valence and conduction bands, which come to exist along any wave vector direction, appear under the second configuration in Fig. 3(f). There are five/three energy subbands, with weak oscillatory dispersion relations near the $\Gamma$ point/ the K and M points. They are responsible for the formation of partially flat bands and thus the high density-of-states near the Fermi level [discussed later in Fig. 5(f)]. The third case in Fig. 3(g) shows a gapless Dirac cone structure at the K/K$^\prime$ valley only with a single Fermi momentum state [a zero density-of-states at $E_F$ in Fig. 5(g)]]. There are more finite energy spacings between valence and conduction subbands at the $\Gamma$ valley. For any guest-atom configurations, most electronic energy spectra are dominated by the silicon atoms, while the opposite is true for the carbon atoms.


\begin{figure}
\centering
\includegraphics[width=10cm, height=18cm]{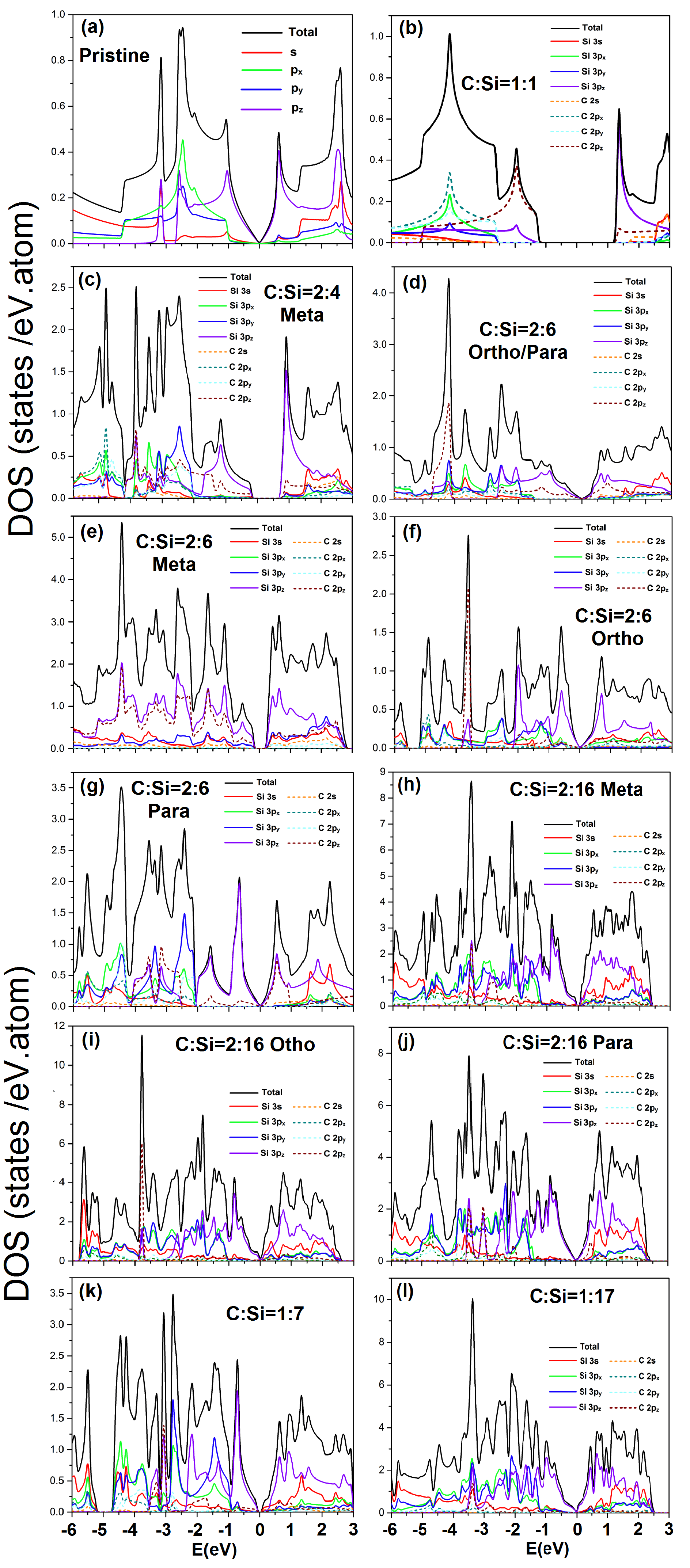}
\caption{ The various van Hove singularities in atom- and orbital-decomposed density-of-states for carbon-substituted silicene systems under the various cases: (a) pristine, (b) [1:1]-, (c) [2:4]-meta-, (d) [2:4]-ortho-/para-, (e) [2:6]-meta-, (f) [2;6]-ortho-,  (g) [2:6]-para-, (h) [2:16]-meta-, (i) [2:16]-ortho-, (j) [2:16]-para-, (k) [1:7]-single-, and (l) [1:17]-single-configurations, being, respectively, illustrated by the solid/dashed [red, green, blue and purple] curves for the [3s, 3p$_x$, 3p$_y$, 3p$_z$] orbitals/[2s, 2p$_x$, 2p$_y$, 2p$_z$] contributions.}
\label{fgr:6}
\end{figure}

\medskip
\par
Clearly, there appear more complex electronic energy spectra when the guest-atom concentrations are decreased [under the reduced/enhanced number of Si-C and C-C/Si-Si bonds]. Typically, three sorts of distribution configurations are revealed under the specific [2:16] condition, the meta-, ortho-, and para-ones in Figs. 3(h) through 3(j). Their low-lying band structures exhibit strongly modified Dirac cones, respectively, with a direct gap of ${E_g\sim\,0.14}$eV due to parabolic valence and conduction dispersion relations near the $\Gamma$ valley, the gapless and crossing structure [the finite energy spacing] along the $\Gamma$M direction [the $\Gamma$K direction],   and   similar zero-gap behavior. Additionally, the single [1:7] substitution in Fig.  3(l) creates an ${E_g\sim\,0.09}$ eV indirect band gap associated with the valence K state and conduction $\Gamma$ state. It should be emphasized that the second and third electronic structures only belong to zero-gap semiconductors [the vanishing density-of-states at $E_F$ in Figs. 5(i) and 5(j)]. The Moir\'e superlattices  have created a substantial number of energy subbands within a smaller first Brillouin zones, leading to frequent band crossings and anti-crossings. As a consequence, it is very difficult to characterize the $\pi$-band width [similarly for the $\sigma$ bandwidths. However, the orbital-projected density-of-states might be useful in examining the relationship among the $\pi$, $\sigma$ and sp$^3$ bondings.  Interestingly,  C-substituted silicene will gradually recover to the pristine case [Fig. 3(a)] at very low guest-atom concentrations, e.g., the appearance of an almost gapless Dirac cone under the [1:17] case in Fig. 3(l). In summary, the energy subbands, with ${|E^{c,v}<1.5|}$ eV, are dominated by the [2p$_z$, 3p$_z$] orbital hybridization, and the other deeper-/higher-energy electronic states  might be closely related to the [2p$_x$, 2p$_y$, 2s, 3p$_x$, 3p$_y$ 3s] or eight orbitals in Si and C atoms. The carbon substitutions in silicene can only induce the modified Dirac cone structures, but not free carriers [the $n$- or $p$-type dopings].

\medskip
\par
 High-resolution angle-resolved photoemission spectroscopy (ARPES) measurements provide a convenient way for examining/identifying occupied electronic energy spectra along specific 2D/1D wave vector directions. This technique has identified  several unusual low-lying energy dispersion relations in graphene-related sp$^2$-bonding materials, such as, the linear valence Dirac cone  in monolayer graphene \cite{9-61}, two pairs of parabolic bands in bilayer AB stacking \cite{9-62}, partially flat, sombrero-shape  and linear energy bands in trilayer ABC stacking \cite{9-63},  bilayer-/monolayer-like energy bands of Bernal graphite  at the K/H corners [AB-stacked one at ${k_z=0}$/${k_z=\pi}$; \cite{9-64}, and 1D parabolic energy subbands of graphene nanoribbons \cite{9-65}, The delicate and thorough  ARPES examinations are required for the predicted band structures of C-substituted silicene materials. That is, they cover the zero- and finite-gap behaviors with the modified Dirac-cone structures across  the Fermi level/without the carrier dopings, the $\pi$ band/subbands with the initial K or $\Gamma$ valley, the intermediate saddle-M-point form $\&$ the final $\Gamma$ valley, the degenerate $\sigma$ bands initiated at deeper energies from the $\Gamma$ valley, their band crossings $\&$ anti-crossings along the K$\Gamma$ and M$\Gamma$ directions, and spin-degenerate electronic states, being very sensitive to the configuration and concentration of guest atoms [Figs. 3(a) through 3(l)]. The full information about the verified band structures is very useful in understanding the substitution-enriched relations among the $\pi$, $\sigma$ and sp$^3$ bondings in Si-Si, Si-C and C-C chemical bonds, and the non-magnetic property.

\medskip
\par

The first-principles calculations acknowledge that the tight-binding model is capable of simulating the low-lying spin-degenerate energy dispersions in carbon-substituted silicene systems. The site energies of different orbitals, the impure/pure multi-orbital hybridizations, the non-uniform environments [the optimal positions of host and guest atoms], and the buckled/planar structures need to be taken into account simultaneously. For example, the full C-substitution, which covers the uniform Si-C bonds in the smallest unit cell, possesses the distinct orbital energies of [3s, 3p$_x$, 2s, 2p$_x$] and the impure $\pi$ [(3p$_z$-2p$_z$)] and $\sigma$ [(3s, 3p$_x$, 3p$_y$)-(2s, 2p$_x$, 2p$_y$)] chemical bondings. In general, it is difficult to determine the various hopping integrals due to the multi-orbital hybridizations, especially for those of Moir\'e superlattices with many atoms in large unit cells. Maybe,   Harrison's rule \cite{9-66} is  an efficient method for solving the non-uniform chemical bonds under dilute substitutions. When the suitable parameters are achieved under numerical fitting with the first priciples calculations, the generalized tight-binding models in the presence of external electric and magnetic fields\cite{9-67,9-68} are very powerful  in completely  understanding the rich magnetic quantization phenomena of Si-C-based 2D compounds. The similar simulation methods could be done for the more complicated B- and N-substituted silicene compounds.

\subsection{Spatial charge density distributions}
\medskip
\par

 The spatial charge density distributions are able to provide certain observable evidences regarding the existence of the impure/pure $\pi$ and $\sigma$ chemical bondings as well as their non-orthogonalities or orthogonalities. Pristine 2D silicene, as clearly displayed in Fig. 4(a) with the ${[x, z]}$- and ${[x, y]}$-plane projections, shows the well-defined ${3p_z}$-${3p_z}$ and ${[3s, 3p_x, 3p_y]}$-${[3s, 3p_x, 3p_y]}$ orbital hybridizations, especially for the latter. In our notation, $\rho$ is a symmetric distribution around the Si-Si bond center. Most of the charge density is accumulated between two silicon atoms [the red region], corresponding to the very strong $\sigma$ bonding of three orbitals. The neighboring Si-atoms are also attracted together through the parallel  ${3p_z}$ orbitals perpendicular to the ${[x, y]}$ plane, in which the $\pi$ bonding appears in the outer region shown by the light green/blue color. As a result of buckling, they have the weak, but significant ${sp^3}$ hybridizations under the non-orthogonality of $\pi$ and $\sigma$ chemical bondings. When the Si-Si bonds become Si-C ones under full substitution, $\rho$ presents a dramatic transformation, being clearly illustrated in Fig. 4(b). The charge density is highly asymmetric with respect to the C-Si bond center, mainly owing to the different electron affinities of the guest and host atoms. There exist more carriers around the guest C-atoms [the red region]; that is, electrons are transferred from silicon to carbon atoms. The whole C-Si bonds consist of a planar honeycomb lattice with a stronger $\sigma$ bonding, compared with those of a pristine one [Fig. 4(a)]. The $\sigma$ and $\pi$ bondings could be roughly defined under the impure ${2p_z}$-${3p_z}$ and ${[2s, 2p_x, 2p_y]}$-${[3s, 3p_x, 3p_y]}$ orbital hybridizations, respectively. They are directly reflected in the $\pi$- and $\sigma$-electronic valence subbands, with direct crossings [Fig. 3(b)]. Most important, two kinds of chemical bondings are orthogonal to each other. Therefore, the ${sp^3}$ orbital hybridizations are negligible in the [1:1] case. Each Si-C bond has the identical chemical environment, i. e., only one chemical bond in a unit cell. This clearly illustrates that it is relatively easy to simulate the first-principles band structure using the tight-binding model \cite{9-69}.

\medskip
\par

 When the carbon concentrations are decreased [Figs. 4(c) through 4(h)], the existence of $\pi$ and $\sigma$ chemical bondings is relatively easily examined from the spatial charge densities. There exist C-Si, Si-Si and even C-C bonds, for which the last ones are stable under the specific ortho cases [Table 1]. The almost symmetric carrier distributions are clearly revealed between two silicon atoms for any concentration and configuration, in which they possess the lowest charge density among three kinds of chemical bonds. Similar phenomena appear for C-C bonds with the highest carrier densities, e.g., $\rho$ values in Figs. 4(d) and 4(f) at ${[2:4]}$  and ${[2:6]}$ ortho conditions, respectively. Roughly speaking, the spatial charge densities in different chemical bonds are not sensitive to changes in various carbon substitutions. These further illustrates that the $\sigma$ and $\pi$ chemical bondings might be well separated from each other. Therefore, they could be roughly defined in the carbon-substituted silicene systems. As a result, the fundamental low-energy  properties are expected to be dominated by the $\pi$-electronic states due to the modified Dirac cone structures. On the other hand, the different chemical bonds lead to complex orbital hybridizations and thus contribute to the difficulties in  obtaining suitable phenomelogical models.  When the first principles electronic energy spectra along the high-symmetry paths are successfully simulated by the tight-binding model with the non-uniform and multi-/single-orbital hopping integrals \cite{9-70}, the diversified essential properties could be fully explored in the near-future, e.g., the rich and unique magnetic quantization phenomena \cite{9-68}, as predicted/observed in layered graphene systems \cite{9-71}.

\medskip
\par

\subsection{Density-of-states}

   There are four/five categories of van Hove singularities in carbon-substituted  silicene systems or pristine one, as clearly illustrated in Figs. 5(a) through 5(f). The orbital- and atom-decomposed density-of-states are very useful in fully understanding the bonding-induced special structures. The critical points, i.e., the band-edge states, in the energy-wave vector space include the linear Dirac cone structure, local minima or maxima of parabolic energy dispersion, saddle points, constant energy loops,  which are closely related to band anticrossings, and partially flat bands [Figs. 3(a) through 3(l)]. Their densities-of-states  generate the V-shape form, discontinuous shoulders, logarithmic divergent peaks, asymmetric peaks in the square root divergence, and delta-function-like peaks, respectively. A pure monolayer silicene in Fig. 5(a) displays an almost linear  $E$-dependence across the Fermi energy with a vanishing density-of-states [a quasi-V shape in the range  ${-0.5}$ meV${\le\,E\le0.5}$ meV; the purple curve of 3p$_z$ orbitals], a logarithmic symmetric $\pi$-peak/${\pi^\ast}$-peak at $-$1.02 eV/0.59 eV, the [$\pi$, $\sigma$]-mixing created square root asymmetric peaks at [$-$2.20 eV, $-$2.65 eV, $-$3.15 eV. The red, green and blue curves of (3p$_x$, 3p$_x$, 3p$_y$) orbitals], the $\Gamma$-valley $\pi$ shoulder at $-$3.20 eV, the initial $\sigma$-[3p$_x$, 3p$_y$] shoulder at ${-1.03}$ eV at the $\Gamma$ point, their M saddle point symmetric peak at ${-2.47}$ eV and the K valley shoulder at ${-4.40}$ eV.   Consequently, the $\pi$  and $\sigma$
bandwidths are $\sim$3.15 eV and 3.36 eV, respectively, It should be noted that the 3s orbitals [the red curve] are frequently accompanied by [3p$_x$, 3p$_y$] ones, but their contributions become significant at the deeper/higher energies, e.g., density-of-states within ${E<-3}$ eV.  The above mentioed features of van Hove singularities further illustrate the well behaved $\pi$ and $\sigma$ chemical bondings and their weak, but important hybridization.

\medskip
\par

 The structural features, energy and number of van Hove singularities exhibit a dramatic transformation under full carbon substitution, as indicated in Fig. 5(b). The special structures cover   vanishing density-of-states within a band gap of ${E_g\sim\,2.56}$ eV at the K/K$^\prime$ valley, the initial $\pi$ shoulder/delta function-like $\pi^\ast$ peak at $-$1.26 eV/1.26 eV [also shows the low-lying $\pi$ and $\pi^\ast$ bands in Fig. 3(b)], the $-$1.95 eV symmetric $\pi$ peak in the logarithmic divergence, the final $\pi$ shoulder at $-$5.00 eV, the first $\sigma$ shoulder at $-$2.75 eV, the $-$4.10 eV logarithmic peak, and the second discontinuous structure  at $-$7.01 eV. We also, observe  that the first valence/conduction  structure is dominated by the C-2p$_z$ orbitals/Si-3p$_z$ ones due to an obvious difference of ionization energy. The widths of the $\pi$ and first $\sigma$ bands are    $\sim$3.74 eV and  4.26 eV, respectively. Additionally, there is no evidence of $\pi$-$\sigma$ band mixing. This is mainly due to the absence of the simultaneous four-orbital structures. The fact that the sp$^s$ bonding is absent agrees with the direct $\pi$-$\sigma$ subband crossings  [Figs. 3(b)].

\medskip
\par

 Within the whole energy range, the van Hove singularities  become more complex during the decrease of guest-C-atom concentration, as a result of zone-folding effects as well as the significant [3s, 3p$_x$, 3p$_y$, 3p$_z$]-[2s, 2p$_x$, 2p$_y$, 2p$_z$] multi-orbital hybridizations. For example, the [2;4]-meta and [2:4]-ortho configurations, as clearly illustrated in Figs. 5(c) and 5(d), exhibit   diverse low-energy van Hove singularities  arising from the dominant 3p$_z$-2p$_z$ chemical bonding. The former has a $\pi$-electronic zero density-of-states in the 0.90 eV-gap region [the $\Gamma$ valley in Fig. 3(c)], a threshold valence shoulder at ${-0.45}$ eV [the first composite conduction state structure due to the discontinuous shoulder and delta function-like peak at 0.45 eV], a strong logarithmic peak at ${-1.41}$ eV, as well as the second  and third step structures at ${-1.69}$ eV  and ${-2.71}$ eV, respectively. Regarding the latter, the 3p$_z$-2p$_z$-diversified van Hove singularities show a gapless V-shape  [a Dirac cone energy spectrum in Fig. 3(d)], the first valence shoulder at ${-0.5}$ eV [the initial conduction shoulder at 0.30 eV], the  logarithm-step composite structure at ${-1.01}$ eV, the symmetric peak arising from two opposite shoulders at ${-1.32}$ eV, a similar one at ${-2.10}$ eV, and the ${-2.65}$-eV asymmetric peak in the square root form. The widths  of the $\pi$-band in the former and latter cases are estimated to be $\sim$4.05 eV and 4.70 eV, respectively. Furthermore, the first $\sigma$-bands  are wider than 4.5 eV. Obviously, the low-energy physical phenomena are dominated by the 3p$_z$ orbitals of Si-host atoms.

\medskip
\par

 As the guest-atom concentration is declined, there appear additional van Hove singularities. Overall, in this way, one is able to characterize the width of the $\pi$-band from the effective distributions of the [3p$_z$, 2p$_z$] orbitals for any concentration and configuration [Figs. 5(a) through 5(l)]. However, the opposite is true for the first $\sigma$-band width except for full substitution and the pristine cases [Figs. 5(a) and 5(b)]. Certain van Hove singularities, close to the Fermi level, are very useful to comprehend the band-edge states in the first pair of the valence and conduction bands. The main features cover the vanishing density-of-states at/across the Fermi level [Figs. 5(c) through 5(e)]/[Figs. 5(d) through 5(g)], the electron-hole asymmetry near the first pair of clearly identifiable shoulders, the prominent $\pi$ peak in logarithmic form, the second and third shoulders and so on. Finally, the dip structure at $E_F$, with a very small gap comparable to the broadening factor, will come to exist for a sufficiently low concentration, e.g., the almost gapless behavior in the [1:17] case [Fig. 5(l)]. We also noticed that the single-particle interactions of spin-orbital couplings cannot create any significant effects on the electronic properties and thus the other essential properties.

\medskip
\par

 High-resolution scanning tunneling spectroscopy (STS) is an efficient method which could serve to identify the valence and conduction band-edge states [the van Hove singularities of the density-of-states. So far, they have confirmed the rich and unique density-of-states in dimension, stacking, layer number dependent graphene based materials, such as, the width and edge structure dominated/chirality and radius related band gaps and considerable asymmetric prominent peaks in 1D semiconducting graphene nanoribbons/carbon nanotubes \cite{9-72}/\cite{9-73}, the V-shape structure with a dip at the Fermi level in monolayer graphene \cite{9-74}, the Dirac point red shift thorough alkali adatom chemisorption [$n$-type doping] \cite{9-75}, the gate-voltage-created band gap in bilayer AB-stacked graphene \cite{9-76}, surface state peak just at $E_F$ in trilayer ABC stacked graphene \cite{9-77}, the semimetallic behavior [finite density-of-states at $E_F$] and the clearly symmetric $\pi$ and ${\pi^\ast}$ peaks at mid energies for Bernal graphite \cite{9-78}.  Similar STS examinations could be conducted on the various van Hove singularities in  carbon-substituted silicene systems, covering the zero, finite gap $\&$ semimetallic band properties, the $\pi$ and $\sigma$ band widths, and the configuration as well as concentration dependent five types of special  structures [V shape across the Fermi level, shoulders, symmetric peaks in the logarithmic form, delta function-like ones, i.e.,  square root asymmetric peaks in Figs. 5(a) through 5(l)]. Both experimental STS and ARPES results would provide sufficient information on the strength relations among the $\pi$, $\sigma$ and sp$^3$ bondings during the chemical substitutions of guest atoms.

\vskip 0.6 truecm
\par\noindent
{\bf 4. Concluding Remarks and Summary }
\vskip 0.3 truecm


\medskip
\par

Chemisorption and substitution of Si-guest atoms on monolayer graphene present unusual geometric properties, which are directly reflected in the spatial charge distributions. Si-adsorbed graphene is a non-buckled honeycomb lattice under  optimal bridge-site positioning. A planar structure clearly indicates a very small variation in the $\sigma$ bonding of C-[2s, 2p$_x$, 2p$_y$] orbitals. Therefore, they hardly take part in Si-C bonds. This adsorption configuration is similar to that of graphene oxide \cite{9-125}, but there is a slight buckling in the latter. The Si-C bond length is about 2.1-2.5 $\AA$, leading to the significant multi-orbital hybridizations. Furthermore, the C-C bond lengths remain almost unchanged. According to the detailed analyses on the 3D charge density distributions, their spatial variations, as well as the atom and orbital projected density-of-states, the sp$^3$-p multi-orbital hybridizations are deduced to determine the chemical Si-C bonds.

\medskip
\par

The theoretical  predictions on the bridge site and Si-C/C-C bond lengths could be examined using high-resolution STM/TEM/LEED measurements.  Also, Si-substituted graphene systems display planar geometries under multi-orbital chemical bondings in Si-C/C-C/Si-Si bonds. The Si-C bond length is about 1.62-1.83 $\AA$, being shorter than those in the Si-adsorbed systems. This means that four orbitals in host and guest atoms are strongly hybridized with one another. The STM experiments are suitable for examining the predicted Si-C, C-C and Si-Si bond lengths. Any chemisorption and substitution of Si-guest atoms significantly modify the unusual band structure of monolayer graphene, especially the zero gap semiconducting behavior and linear Dirac cone due to the $\pi$ bondings of the 2p$_z$ orbitals. In the 100$\%$ double- and single-side adsorption cases, Si-adsorbed graphene systems are semimetals with free conduction electrons and valence holes. For the low concentration systems, they belong to the $p$-type metals with only free valence holes. In general, the Dirac cone structure near the stable valley is seriously distorted after Si-chemisorption.  There are more valence and conduction energy subbands, accompanied by the various band-edge states within an entire  energy spectrum, e.g., the emergent low-lying energy bands along the K$\Gamma$ and M$\Gamma$ directions. However, the $\sigma$ bands, which arises from the [2p$_x$, 2p$_y$] orbitals of carbon atoms, exhibit a robust red shift of $\sim$1 eV and  $\sim$0.5 eV for 100$\%$ and lower-concentration cases, respectively.

\medskip
\par

The above mentioned important results directly reveal critical mechanisms, i.e., , the multi-orbital hybridizations of sp$^3$-p in Si-C bonds, sp$^3$-sp$^3$ in Si-Si bonds, and sp$^2$-sp$^2$/$\pi$ in C-C bonds. Such chemical bondings mainly produce four Si- and C-orbitals, as a result of the bridge-site adsorption positions, as well as the binding energies and charge distributions of the separated orbitals.  High resolution ARPES measurements are very useful in verifying the low-lying valence bands near/crossing the Fermi level along $\Gamma$K and $\Gamma$M and the rigid $\sigma$ bands initiated from the $\Gamma$ valley. On the other hand, all substitution results in semiconducting behavior with a finite or zero band gap. The Dirac cone structure presents a deviation from the $\Gamma$ point, a strong distortion, or even destruction. The number of valence and conduction energy subbands remains the same after chemical substitutions. Furthermore, they are co-dominated by the Si-guest and C-host atoms.

\medskip
\par

The main features of the band structure in Si-substituted graphene are dominated by the sp$^2$-sp$^2$ and  p-p orbital hybridizations in Si-C/C-C/Si-Si bonds. Several unusual van Hove singularities in the atom and orbital decomposed density-of-states are created under the Si-guest-atom chemisorption and substitution. Pristine monolayer graphene only presents a V-shaped structure with a vanishing value at the Fermi level, the logarithmic symmetric $\pi$-/${\pi^\ast}$-peaks at $-$2.41 eV/1.81 eV, and a shoulder structure at ${-3.01}$ eV, in which the former two and the last one, respectively, correspond to the $\pi$ and $\sigma$ bondings of carbon atoms. The strong $\pi$-bonding evidence is  totally destroyed by chemisorption except for the very diluted Si-adatoms. They are replaced by a finite density-of-states at ${E_F= 0}$, many shoulders, and peak structures. Furthermore, they are co-dominated by the four Si-[3s, 3p$_x$, 3p$_y$, 3p$_z$] orbitals and the single C-2p$_z$ orbital, since their contributions are merged together. Specifically, the $\sigma$-band shoulder comes to exist at $-$4.21 eV, $-$4.11 eV or $-$3.5 eV, and the $\sim$0.5-1.0 the red shift is closely related to the distinct ionization energies of Si and C atoms. The above-mentioned significant features further support and illustrate the sp$^3$-sp$^3$ $\&$ sp$^2$-sp$^2$ multi-orbital hybridizations in Si-C bonds and C-C bonds, respectively. Concerning the substitution cases, density-of-states is zero within a finite energy range centered at the Fermi level, except for few systems with deformed V-shaped structures at $E_F$ . All the C-substituted silicene systems are finite- or zero gap semiconductors. The special structures, which originate from Si-3p$_z$ and C-2p$_z$ orbitals, appear simultaneously. Also, a similar behavior is revealed in Si-[3p$_x$, 3p$_y$] and C-[2p$_x$, 2p$_y$] orbitals. These further illustrate the critical single- and multi-orbital hybridizations in Si-C bonds. The predicted van Hove singularities could be verified by the high-resolution STS experiments \cite{9-126}.

\par\noindent {\bf Acknowledgments}

This work was financially supported by the Hierarchical Green-Energy Materials (Hi-GEM) Research Center, from The Featured Areas Research Center Program within the framework of the Higher Education Sprout Project by the Ministry of Education (MOE) and the Ministry of Science and Technology (MOST 108-3017-F-006 -003) in Taiwan.

\newpage
\renewcommand{\baselinestretch}{0.2}

 \newpage
      \centerline {\Large \textbf {Table Captions}}

     \bigskip \vskip0.6 truecm

           Table 9.1: The optimal geometric structures of the carbon-substituted silicene systems under the various concentrations and configurations with the C-C, C-Si $\&$ Si-Si bond lengths, the band gaps, the ground state energies  per unit cell, and the height differences between A and B sublattices.

\end{document}